\documentclass[twocolumn]{article}

\usepackage{graphicx}

\newcommand{\partderiv}[2]{\frac{\partial #1}{\partial #2}}

\newcommand{\calM}{{\cal M}}
\newcommand{\calX}{{\cal X}}
\newcommand{\calN}{{\cal N}}
\newcommand{\calB}{{\cal B}}

\title{Energy conservation and equivalence principle\\ 
in General Relativity}
\author{Michael B. Mensky\\
{\small P.N.Lebedev Physical Institute,} 
{\small 53 Leninsky prosp., 117924 Moscow, Russia}}
\date{June 18, 2004}
 
\begin{document}

\maketitle

\begin{abstract}
The generalized Stokes theorem (connecting integrals of dimensions 3 and 4) is formulated in a curved space-time in terms of paths in Minkowski space. A covariant integral form of the conservation law for the energy-momentum of matter is then derived in General Relativity. It generalizes Einstein's equivalence principle, excluding gravitational field from the conservation law for the energy-momentum of matter. 
\end{abstract}

%\tableofcontents 

\section{Introduction}
\label{sec:Intro}

The question of conserving energy-momentum in General Relativity (GR) always attracted much attention. One of the reasons is that covariant description of energy-momentum seems to be incompatible with the integral conservation law. Particularly, it is generally believed that no integral conservation law follows from the covariant differential conservation law for the energy-momentum tensor (EMT) of matter (i.e. from its covariant divergence being zero) \cite{EMT,Landau}. 

We shall show that this is not quite right: the nullification of the divergence of material EMT yields a sort of integral conservation law, although it is formulated in unusual way. The \emph{covariant integral conservation law} for the material EMT will be formulated in terms of path groups introduced earlier for the alternative formulation of gauge theory and gravity (see \cite{PathGroup03} and references therein). The formalism of path groups is a group-theoretical version of the Mandelstam's path-dependent wave functions formalism and the Yang's non-integrable phase factors \cite{MandelstamYang}. 

The resulting form of the conservation law for the material EMT is unusual. It is formulated not directly in the physical space-time $\calX$ but rather in a ``standard flat space-time''. The latter is the Minkowski space $\calM$ which may be considered as a model of any tangent space to $\calX$. The standard space-time $\calM$ is connected with the physical space $\calX$ in a non-holonomical way. This is used for characterization of points in $\calX$ in terms of paths in $\calM$. 

It has been shown earlier that the equivalence principle (EP) for the mechanical law of motion as well as its quantum generalization may be naturally formulated in terms of the `standard' space-time $\calM$ \cite{PGequivPrinciple}. The material energy-momentum conservation law will be formulated also in terms of $\calM$ and interpreted as EP for energy: gravitational field may be excluded from the formulation of this conservation law. 

\quad

Let us set forth the problem systematically. In the flat (Minkowski) space-time, i.e. in the case of null gravity, energy and momentum are described locally by the corresponding EMT $T_{\alpha\beta}$ ($\alpha,\beta = 0,1,2,3$) which satisfies the `differential conservation law' 
\begin{equation}
	\partial_\beta T^{\alpha\beta} = 0
	\label{NullDivergNoncovar}
\end{equation}
where $\partial_\beta = \partial/\partial \xi^\beta$ are derivatives in the coordinates $\xi^\beta$ of Minkowski space ($\beta = 0,1,2,3$). 

The corresponding integral conservation law follows (see for example \cite{Landau}) from the 4-dimensional Gauss theorem 
\begin{equation}
	\int_{\partial W} T^{\alpha\beta} \, d\sigma_\beta 
	= \int_W \partial_\beta T^{\alpha\beta} \, dw
	\label{GaussFourDimNoncovar}
\end{equation}
where $W$ is a 4-dimensional region of space-time, $\partial W$ its (3-dimensional) boundary, $dw$ and $d\sigma_\alpha$ are correspondingly elements of 4-volume and of area of the 3-dimensional hypersurface in Minkowski space. It follows from Eq.~(\ref{NullDivergNoncovar}) that the integral in the r.h.s. of Eq.~(\ref{GaussFourDimNoncovar}) is zero for any $W$. Therefore, the l.h.s. is null for the boundary $\partial W$ of an arbitrary 4-volume. 

Let $W$ be a 4-volume swept by some 3-volume $V$ during a time interval $[t_1, t_2]$. Then the boundary $\partial W$ of $W$ consists of two space-like hypersurfaces (`bottom' and `top' ones) presenting the 3-volume $V$ at the two time moments $t_1$ and $t_2$, and a time-like `side' hypersurface which presents the boundary $\partial V$ of $V$ in all times between $t_1$ and $t_2$. The null value of the l.h.s. of (\ref{GaussFourDimNoncovar}) is then naturally interpreted as an integral form of the conservation law: the difference  between energy-momentum in the volume $V$ at times $t_2$ and $t_1$ is equal to the flow of energy-momentum throw the boundary of $V$ during the interval $[t_1, t_2]$. 

Eq.~(\ref{GaussFourDimNoncovar}) is a special case of the generalized Gauss theorem 
\begin{equation}
		\int_{\partial W} H	= \int_W dH
	\label{GenStokesNoncovar}
\end{equation}
where $H = H_{\alpha\beta\gamma}\, d\xi^{\alpha}\wedge d\xi^\beta\wedge d\xi^\gamma$ is 3-dimensional \emph{external form} and $dH = \partial_{[\alpha}H_{\beta\gamma\delta]}\, d\xi^{\alpha}\wedge d\xi^\beta\wedge d\xi^\gamma\wedge d\xi^\delta$ its (4-dimensional) \emph{external derivative} (the sign $[\, ]$ here stands for antisymmetrization). 

The 4-dimensional Gauss theorem  (\ref{GaussFourDimNoncovar}) is obtained from Eq.~(\ref{GenStokesNoncovar}) for the four 3-forms $H^{(\alpha')}_{\beta\gamma\delta} = T^{\alpha'\alpha}\, \epsilon_{\alpha\beta\gamma\delta}$ ($\alpha' = 0,1,2,3$) if it is denoted: 
\begin{eqnarray}
 d\xi^0\wedge d\xi^1\wedge d\xi^2\wedge d\xi^3 &=& \delta  w, 
\nonumber\\ 
 d\xi^\alpha\wedge d\xi^\beta\wedge d\xi^\gamma 
&=& \epsilon^{\alpha''\alpha\beta\gamma} \, \delta \sigma_{\alpha''}.
\end{eqnarray}
Here $\epsilon^{\alpha\beta\gamma\delta}$ is the 4-dimensional completely antisymmetric \emph{discriminant tensor} defined by $\epsilon^{0123}=1$. 

Let us make the Stokes theorem more concrete by parametrizing the manifolds of integration as 
\begin{eqnarray}
W: \quad	
	\xi &=& \xi(\tau_1, \tau_2, \tau_3, \tau_4), \quad 
	0\le\tau_i\le 1, \nonumber\\
\partial W: \quad		
	\xi &=& \widetilde\xi(\tau_1, \tau_2, \tau_3) 
= \xi(\tau_1, \tau_2, \tau_3, 1). 
\end{eqnarray}
Then the Stokes theorem (\ref{GenStokesNoncovar}) connecting the $3D$ and $4D$ integrals may be rewritten as 
\begin{equation}
	I_H(\partial W) = I_{dH}(W) 
	\label{NonCovarStokesXi}
\end{equation}
where
\begin{eqnarray}
			I_H(\partial W)
&=& 3! \int_0^1 d\tau_1  \int_0^1 d\tau_2 \int_0^1 d\tau_3  \nonumber \\
	&\times& 
	\partderiv{\widetilde\xi^{\alpha}}{\tau_1}
	\cdot \partderiv{\widetilde\xi^{\beta}}{\tau_2}
	\cdot \partderiv{\widetilde\xi^{\gamma}}{\tau_3}\;
	H_{\alpha\beta\gamma}(\widetilde\xi),
	\label{NonCovarStokesIntegrals}\\
I_{dH}(W) 
	&=& 4! \int_0^1 d\tau_1  \int_0^1 d\tau_2 
	\int_0^1 d\tau_3 \int_0^1 d\tau_4 \nonumber \\
	&\times& 
	\partderiv{\xi^{\alpha}}{\tau_1}
	\cdot \partderiv{\xi^{\beta}}{\tau_2}
	\cdot \partderiv{\xi^{\gamma}}{\tau_3}
	\cdot \partderiv{\xi^{\delta}}{\tau_4}\;
	\partial_{[\alpha}
	H_{\beta\gamma\delta\, ]}(\xi) .	\nonumber 
\end{eqnarray}

\quad

Consider now special features of EMT in General Relativity (GR) which is formulated in a curved (pseudo-Riemannian) space-time $\calX$. Denote the coordinates in $\calX$ as $x^\mu$, and the derivatives in these coordinates as $\partial_\mu = \partial/\partial x^\mu$, $\mu = 0,1,2,3$. If the divergence $\partial_\nu T^{\mu\nu}$ of EMT had everywhere null value, the conservation of energy-momentum could be derived in the usual way (see above). This however is not the case in GR. 

Einstein equations of GR have the form 
\begin{equation}
	R_{\mu\nu} - \frac 12 \, g_{\mu\nu} R = \kappa T_{\mu\nu},
	\label{EinstEq}
\end{equation}
where the r.h.s. is proportional to the EMT of matter $T_{\mu\nu}$ and the l.h.s. is Einstein tensor. It follows from the so-called Bianchi identity that the \emph{covariant divergence} (expressed through the covariant derivatives $\nabla_\mu$ instead of $\partial_\mu$) of Einstein tensor is zero. Therefore, the same is valid for the EMT of matter:
\begin{equation}
	\nabla_\nu T^{\mu\nu} = 0.
	\label{CovarDiffConserv}
\end{equation}

No integral energy-momentum conservation law could be derived from the covariant equation (\ref{CovarDiffConserv}) with the help of the usual procedure sketched above. To solve the problem, Einstein introduced a pseudotensor presenting energy-momentum of gravitational field. One may (see for example \cite{Landau}) rewrite the covariant differential law (\ref{CovarDiffConserv}) in the form 
\begin{equation}
	\partial_\nu (-g)(T^{\mu\nu} + t^{\mu\nu})= 0
	\label{PseudotensorDiverg}
\end{equation}
where the non-tensor entity $t_{\mu\nu}$ (transforming non-covariantly under change of the coordinate frame) is the \emph{pseudotensor of energy-momentum of gravitational field}. Eq.~(\ref{PseudotensorDiverg}) yields the conservation law for the total (material + gravitational) energy-momentum but with the localization of gravitational energy-momentum described by the pseudotensor instead of a tensor, i.e. non-covariantly. 

Equation (\ref{CovarDiffConserv}) for material EMT is called sometimes covariant differential energy-momentum conservation law because of the evident analogy with Eq.~(\ref{NullDivergNoncovar}). However, no integral conservation law has been derived from the covariant equation (\ref{CovarDiffConserv}). 

We shall show that an integral conservation law for the material EMT $T^{\mu\nu}$ does in fact follow from the differential conservation law (\ref{CovarDiffConserv}). The price paid for this feasibility is that the \emph{path-dependent forms} are used in the corresponding integrals instead of the usual local (point-dependent) forms. This reflects the non-holonomic character of the connection between physical space-time $\calX$ and `standard' Minkowski space-time $\calM$. Mathematically this is naturally presented in the framework of the formalism of path groups (see \cite{MenBk83eng,PathGroup03}). 

More concretely, we shall derive analogues of Eqs.~(\ref{NonCovarStokesXi},\ref{NonCovarStokesIntegrals}) and therefore of Eq.~(\ref{GaussFourDimNoncovar}) but for path-dependent forms and covariant derivatives. The resulting integral conservation law includes only EMT of matter (without gravitational field) and may be interpreted as the \emph{extension of EP}: gravitational field may be excluded from the integral energy-momentum conservation law for matter just as it is excluded from the law of motion. 

The key point making this possible is that the covariant derivatives acquire, in the Path Group formalism, a group-theoretical interpretation as generators of elements of the Path Group (generalized translations). Earlier this was used to derive a non-Abelian version of the Stokes theorem for gauge theory \cite{MenBk83eng,PathGroup03}. 

The plan of the paper is following. In Sect.~\ref{sec:PathGroups} the main concepts from theory of path groups will be sketched. In Sect.~\ref{sec:PathGroupGrav} it will be shown how the group of paths in Minkowski space $\calM$ may be applied in a curved space-time $\calX$ (since $\calM$ is identical to the tangent space in any point $x\in\calX$). In Sect.~\ref{sec:EnergyGrav} the $3D/4D$ Stokes theorem for GR will be introduced and in Sect.~\ref{sec:InterpretEnergyMomInteg} the conservation for the material energy-momentum will be formulated in integral form (EP for energy conservation). A short conclusion will be given in Sect.~\ref{sec:Conclusion}. 

\section{Path groups}
\label{sec:PathGroups}

Formalism of path groups (see \cite{MenBk83eng,PathGroup03} and references therein) generalizes the concept of translation onto the case when infinitesimal translations do not commute (as parallel transports in non-Abelian theory or GR do not). Non-commutativity of translations forces us, when composing a finite translation from infinitesimal ones, to specify the order of those. The finite translation is characterized then not only by the difference between the initial and final points, but also by the shape of the curve along which the translation is performed. The resulting (generalized) translation is formalized as a \emph{path}, an element of a path group $p\in P$. Mathematically a path is defined as a class of equivalent continuous curves in the corresponding space, with the appropriate definition of equivalence. 

The most important for us are paths on Minkowski space $\calM$. They form the group $P = P(\calM)$. However paths may also be defined on any (even infinite-dimensional) manifold $G$ provided this manifold has the structure of a group. The group of paths on the group $G$ is denoted by $P(G)$. Since $P = P(\calM)$ is a group, we may define paths on this group. This gives the group $P^{(2)} = P^{(2)}({\cal M}) = P(P({\cal M}))$. Each element of this group is a path on the space of paths on Minkowski space. Such an element will be called \emph{2-path} on $\calM$. It generalizes the concept of (2-dimensional) surface in such a way that this concept be applicable in a non-Abelian theory. A 2-path may be looked on as an \emph{ordered surface}. The group of \emph{$n$-paths} on $\calM$ ($n$-dimensional ordered manifolds) may be defined recurrently as $P^{(n)} = P^{(n)}({\cal M}) = P(P^{(n-1)}({\cal M}))$. We shall need $n$-paths on $\calM$ with $n$ up to $n=4$ to generalize 3-dimensional and 4-dimensional integrals.

In any path group $P = P(G)$ the subgroup of closed paths, or loops, exists, $L\in P$. It contains the curves with the initial and final points coinciding with each other. In particular, in the group of $n$-paths in Minkowski space, the subgroup of $n$-loops, $L^{(n)}(\calM)\subset P^{(n)}(\calM)$, is defined. It contains the curves in $P^{(n-1)}(\calM)$ which begin and end by the same $(n-1)$-path. 

For each path $p$ we shall define the \emph{shift along the path}, $\Delta p$, specifying the difference between its final and initial points. The shift is defined by the subtraction $\Delta p = \xi_{\mathrm{fin}} - \xi_{\mathrm{init}}$ for paths in a linear (e.g. Minkowski) space or by left shift $\Delta p = [g_{\mathrm{init}}]^{-1}\, g_{\mathrm{fin}}$ for paths on a group space. Closed paths have null shifts: $\Delta l = 0$ if $l$ is a loop in a linear space (e.g. $l\in L(\calM)$) or $\Delta l = 1\in G$ for a loop in the group space ($l\in L(G)$). 

Besides `free' paths $p$ forming a group $P$ we shall consider also \emph{pinned paths} $p_\xi$ of the same shape but with fixed initial points. They form a groupoid $\widehat P$ (not any pair of them may be multiplied). Relation between both types of paths is analogous to the relation between free and pinned vectors. 

\quad

Not going into much details, let us comment on some essential points of the definition of the path groups, illustrating them in case of paths in Minkowski space. The main point is that a path is not an individual curve but a class of equivalent continuous curves. The equivalence is defined in such a way that the paths form a group. 

1)~The multiplication $\{\xi''\} = \{\xi\} \{\xi'\}$ of continuous curves $\{\xi(\tau)|0\le\tau\le 1\}$ and $\{\xi'(\tau)|0\le\tau\le 1\}$ (in any manifold, for example in Minkowski space) is defined as the passage along one of them after the other:\footnote{This definition differs from one accepted in the previous papers in that the multiplied paths are ordered from the left to the right (which is convenient for the application in GR).}
\begin{equation}
	\xi''(\tau) = \left\{ 
\begin{array}{lll}
\xi(2\tau) & \mbox{for} & 0\le\tau\le \frac 12\\
\xi'(2\tau - 1) 
& \mbox{for} & \frac 12\le\tau\le 1
\end{array}
\right.
\end{equation}
($\xi'(0) = \xi(1)$ is necessary to obtain again a continuous curve). For this multiplication being associative (as is necessary for a group), each path is characterized by the corresponding curve independently of its parametrization. This means that the curves are considered equivalent if they differ only by the way of parametrization. 

2) The inversion of a curve $\{\xi(\tau)|0\le\tau\le 1\}^{-1}$ is defined as passing the same curve in the opposite direction: $\{\xi'(\tau) = \xi(1-\tau)|0\le\tau\le 1\}$. The multiplication of a path by the inverse path has to give the group unity. This means that a curve obtained by passing along some set of points and just after this along the same set of points in the opposite direction, is equivalent to the constant (null) curve. The class of all such curves is interpreted as the unity of the path group. To make this more manifest, one may think that the backward passage of a curve cancels this curve. The inclusion of such to-and-back-passed curves in the middle of any curve (as an appendage) converts the curve in an equivalent one. 

3) If only these two elements of equivalence are used in constructing the equivalence classes, then the resulting classes form the set $\widehat P$ of \emph{pinned paths}. The initial and final points of each pinned path are fixed. The pinned paths are convenient as a technical instrument. In the present paper we shall formulate the definitions of integrals in terms of pinned paths. However $\widehat P$ is not a group but only a \emph{groupoid}. This means that not any pair of pinned paths may be multiplied: for this being possible, the initial point of one of them has to coincide with the final point of the other. To obtain a group, we have to go over to the \emph{free paths} or simply \emph{paths}. For this end we need one more element of equivalence: two curves are considered equivalent if they differ by a general shift, $\xi'(\tau) = \xi(\tau) + a$ (by a right shift in the case of curves on a group, $g'(\tau) = g(\tau)\,g_0$). Each path $p\in P$ is defined as a class of equivalent curves and may be presented by any of these curves. We shall write in this case $p = \{\xi(\tau)\in\calM\}$ or $p = \{g(\tau)\in G\}$ for paths on a group. If a path $p\in P$ and some point $\xi\in\calM$ (or $g\in G$ in case of the path on a group) are given, the pinned path $p_\xi$ (correspondingly $p_g$) starting in this point is evidently determined. 

\section{Path groups in GR}
\label{sec:PathGroupGrav}

Applying path groups in GR has important special features as compared to their application in gauge theory without gravity. The reason is that the actual (physical) space-time without gravity is Minkowski space $\calM$, hence path groups may be defined on the physical space-time itself: $P = P(\calM)$, $P^{(2)} = P^{(2)}(\calM) = P(P(\calM))$, etc. With non-zero gravitational field, the actual (physical) space-time $\calX$ is pseudo-Riemannian and has generally no group structure. No natural concept of general shift (such as right shift of a group space) can be defined on such $\calX$, and the group of paths on $\calX$ cannot be defined. 

It happens however that the group $P = P(\calM)$ of paths on Minkowski space may do work also in GR. The action of $P$ is defined in this case not in the space-time $\calX$, but in the fiber bundle $\calN$ of orthonormal local frames (bases of tangent spaces) over $\calX$. 

This is possible because a tangent space to any point of $\calX$ has the structure of Minkowski space $\calM$ (note that this is just what underlines Einstein's EP). As a result, a sort of \emph{non-holonomic} correspondence between curves in $\calX$ and paths in $\calM$ (i.e. elements of $P$) may be naturally defined. This correspondence has been used in \cite{PGequivPrinciple} for formulating EP and its quantum generalization in terms of the path group $P(\calM)$. The paths (in $\calM$) associated (in non-holonomic way) with the curves in $\calX$ are called there \emph{flat models} of these curves. EP have an elegant formulation in these terms: the classical particles move (in $\calX$) along the trajectories which have straight lines (in $\calM$) as their flat models. 

Instead of the correspondence between curves in $\calX$ and their flat models in $\calM$ we shall make use of an equivalent machinery: the action of the path group $P = P(\calM)$ on the fiber bundle $\calN$ of orthonormal local frames over $\calX$. Let us define the main concepts of this formalism. Note that the most important for us are the basis vector fields $B_\alpha$ defined by (\ref{BasisField}) and the operators $U(p)$ (given by Eq.~(\ref{Up})) forming a representation of the group $P$. The basis vector fields correspond, in the fiber bundle $\calN$, to the covariant derivatives on the base $\calX$. The operators $U(p)$ describe parallel transports along the curves with the given flat models $p$. 

Let $x^\mu$ ($\mu = 0,1,2,3$) be the coordinates in $\calX$. A \emph{vector field} in $\calX$ may be defined as a differential operator $X = X^\mu \,  \partial/{\partial x^\mu}$ where the components $X^\mu$ (that may depend on the point, $X^\mu = X^\mu(x)$) specify the vector field in the given coordinate system. A tangent vector in the given point $x\in\calX$ may be determined by its components $X^\mu$ in respect to the given coordinates (since $x$ is now given, $X^\mu$ are four numbers rather then a point-dependent field). The set of all vectors in the given point $x$ forms the tangent space $T(x)$ in this point. The set of four linearly independent vectors $\{b_\alpha\in T(x) | \alpha = 0,1,2,3\}$ in the tangent space forms its basis or \emph{local frame} in the point $x$. Each of these vectors, $b_\alpha$, is characterized by their components $b_\alpha^\nu$. Therefore, 20 numbers $\{x^\mu, b_\alpha^\nu | \alpha, \mu, \nu = 0,1,2,3\}$ completely characterize a local frame $b$ (including characterization the point $x$) and may serve as the coordinates in the set $\calB$ of all local frames in all points of $\calX$. The set $\calB$ is a \emph{fiber bundle of local frames} over $\calX$. The space-time $\calX$ is said to be the \emph{base} of the fiber bundle $\calB$. The subset $\calB_x$ of local frames in the given point $x$ is a \emph{fiber} over this point. 

We shall need also the subbundle $\calN\subset\calB$ of all \emph{orthonormal local frames}, i.e. those $n\in\calB$ which satisfy the relation 
$$
g_{\mu\nu} n_\alpha^\mu n_\beta^\nu = \eta_{\alpha\beta}
$$
with $g_{\mu\nu}$ being metric tensor in $\calX$ and $\eta_{\alpha\beta}$ Minkowski tensor (equal to $\mathrm{Diag(1,-1,-1,-1)}$). 

Vector fields on $\calB$ may be given by their components in respect to the coordinates $(x^\mu, b_\alpha^\nu)$ or as a differential operators (cf. vector fields on $\calX$ defined above). An important instrument for us will be so called \emph{basis vector fields} $B_\alpha$, $\alpha = 0,1,2,3$ in $\calB$ closely related to covariant derivatives. They are defined as the differential operators \begin{equation}
	B_\alpha = b_\alpha^\mu \, \partderiv{}{x^\mu} 
	- b_\alpha^\mu b_\beta^\nu \, \Gamma^\lambda_{\; \mu\nu}(x)\,
	\partderiv{}{b^\lambda_\beta}, 
	\label{BasisField}
\end{equation}
where $\Gamma^\lambda_{\; \mu\nu}$ are Christoffel coefficients. The basis vector fields may also be considered as vector fields on $\calN$ (because the displacement along these vector fields does not violate the normalization of local frames). 

The linear combinations $B(a) = a^\alpha B_\alpha$ defined by the Minkowskian 4-vectors $a = \{a^\alpha|\alpha=0,1,2,3\}$ are also called basis vector fields and determine the subspace of \emph{horizontal} vectors in each point of $\calB$ or $\calN$. The curves in $\calB$ or in $\calN$ are called horizontal if they have horizontal tangent vectors in each point. Horizontal curves define \emph{parallel transports} of local frames $b\in\calB$ (or $n\in\calN$) and therefore of tangent vectors $X\in T(x)$ to $\calX$. These operations may be connected with paths in Minkowski space in the following way. 

Let $p\in P = P(\calM)$ be a path in Minkowski space. Associate with this path an operator $U(p)$ defined as an ordered exponential (notice that the ordering is performed from left to right) 
\begin{eqnarray}
	U(p) &=& {\cal P}^\rightarrow 
	\exp\left( \int_p B(d\xi)\right)\nonumber \\
	&=& \lim_{N\to\infty} e^{B(\delta\xi_1)}\dots e^{B(\delta\xi_N)}
	\label{Up}
\end{eqnarray}
where $\xi_i$ ($i=1,\dots,N$) is a `skeletonization' of the path $p$ and $\delta\xi_i = \xi_i  - \xi_{i-1}$. It is important that the operators $U(p)$ form a representation of the group $P$. 

Considering functions on $\calB$ or on $\calN$ and making use of the equalities 
\begin{equation}
	(U(p)\varphi)(b) = \varphi(bp) \quad \mbox{or} \quad
	(U(p)\varphi)(n) = \varphi(np)
	\label{PathActionBundle}
\end{equation}
we can define the action of the path $p$ on these fiber bundles, $p:\; b\rightarrow bp$ and $p:\; n\rightarrow np$. The formulas (\ref{Up}), (\ref{PathActionBundle}) define \emph{parallel transports} along the curves with the given flat model $p$. 

\section{$3D/4D$ Stokes theorem in GR}
\label{sec:EnergyGrav}

Let $H_\mu^\nu(x)$ be a tensor field on $\calX$. Then the corresponding tensor field on $\calN$ (the lifting of the tensor field from the base onto the fiber bundle) is
\begin{equation}
	H_\alpha^\beta(n) 
	= n_\alpha^\mu \, (n^{-1})^\beta_\nu \, H_\mu^\nu(x).
	\label{TensorN}
\end{equation}
Here $n$ is an orthonormal frame in the point $x$ and $n^{-1}$ the inverse matrix for $n = \{n_\alpha^\mu\}$. We shall also define operator $H_\alpha^\beta$ acting on the functions $\varphi(n)$: 
\begin{equation}
	(H_\alpha^\beta\, \varphi)(n) 
	= H_\alpha^\beta(n)\, \varphi(n) 
	\label{TensorOperator}
\end{equation}
The tensor fields on $\calN$ which have arbitrary numbers of lower and upper indices (for example external forms i.e. antisymmetric tensors with lower indices) may be defined in analogous way. 

With the help of the operators $U(p)$ from Eq.~(\ref{Up}) we can also define the \emph{path-dependent tensor} 
\begin{equation}
	{\cal H_\alpha^\beta}(p) = U(p) H_\alpha^\beta U(p^{-1}) 
	\label{PathDependTensor}
\end{equation}
and analogously for tensors of arbitrary ranks. 

Although non-commutative operators $U(p)$ are present in the definition (\ref{PathDependTensor}), the resulting path-dependent tensors commute provided that the corresponding local (point-dependent) ones (\ref{TensorN}), (\ref{TensorOperator}) do. This is readily seen from the following formulas for arbitrary tensors $H_1$ and $H_2$ (we omit their indices): 
\begin{eqnarray}
	&&\Big({\cal H}_1(p_1){\cal H}_2(p_2)\varphi\Big)(n)
	= H_1(n p_1) H_2(n p_2) \varphi(n) 
	\nonumber\\
 &=& \Big({\cal H}_2(p_2){\cal H}_1(p_1)\varphi\Big)(n).
\end{eqnarray}
Hereinafter we shall assume commutativity of the tensors (external forms) in the integrands. This will be important for defining the integrals and for validity of the Stokes theorem. 

\quad

Define a 3-dimensional `path-dependent' integral making use of the concepts elaborated in theory of path groups. An analogue of the concept of (3-dimensional) volume is a 3-path.\footnote{It is enough for us to consider only pinned $n$-paths $\widehat p^{(n)}\in\widehat P^{(n)}$. However, for simplicity we shall use the term `$n$-path' instead of `pinned $n$-path' and notation $p^{(n)}$ instead of $\widehat p^{(n)}$.} Let a 3-path $v$ be defined by the following families of 2-paths and paths: 
\begin{eqnarray}
	v &=& \{ s(\tau_3) | 0\le \tau_3 \le 1\} 
	\in \widehat P^{(3)}\\
s(\tau_3) &=& \{ p(\tau_2,\tau_3) | 0\le \tau_2 \le 1\} 
\in \widehat P^{(2)} \nonumber\\
p(\tau_2,\tau_3) 
	&=& \{ \widetilde\xi(\tau_1,\tau_2,\tau_3) | 0\le \tau_1 \le 1\} \in \widehat P. \nonumber
\end{eqnarray}

Together with an arbitrary path $p\in \widehat P$ we shall need to consider also the paths which coincide with the parts of $p$. If $p = \{ \xi(\sigma') | 0\le\sigma'\le 1\}$, then we shall denote by $p_\sigma$ the part of the path $p$ corresponding to the interval $0\le\sigma'\le\sigma$. In the explicit form $p_\sigma = \{ \xi(\sigma\sigma') | 0\le\sigma'\le 1\}$. 

Let ${\cal H}_{\alpha\beta\gamma}(p)$ be a path-dependent 3-form corresponding to a local 3-form $H_{\mu\nu\kappa}(x)$ and its lifting $H_{\alpha\beta\gamma}(n)$ to $\calN$. Define an integral of this form over the 3-path $v$ as 
\begin{eqnarray}
&&	I_H(v) = 
	3!\int_0^1 d\tau_1 \int_0^1 d\tau_2 \int_0^1 d\tau_3 \; \nonumber\\	&\times&
	\partderiv{\widetilde\xi^\alpha}{\tau_1} \cdot 
	\partderiv{\widetilde\xi^\beta}{\tau_2} \cdot 
	\partderiv{\widetilde\xi^\gamma}{\tau_3} \;
	{\cal H}_{\alpha\beta\gamma}(p_{\tau_1}(\tau_2,\tau_3))
	\label{CovarThreeInt}
\end{eqnarray}
where $\widetilde\xi = \widetilde\xi(\tau_1,\tau_2,\tau_3)$. The function of 3-paths thus defined is additive: $I_H(vv') = I_H(v) + I_H(v')$. 

Analogously, let us introduce the following notations for a 4-path $w$: 
\begin{eqnarray}
	w &=& \{ v(\tau_4) | 0\le \tau_4 \le 1\} 
	\in \widehat P^{(4)}\label{four-path}\\
	v(\tau_4) &=& \{ s(\tau_3,\tau_4) | 0\le \tau_3 \le 1\} \in \widehat P^{(3)}\nonumber\\
s(\tau_3,\tau_4) &=& \{ p(\tau_2,\tau_3,\tau_4) | 0\le \tau_2 \le 1\} \in P^{(2)} \nonumber\\
p(\tau_2,\tau_3,\tau_4) 
	&=& \{ \xi(\tau_1,\tau_2,\tau_3,\tau_4) | 0\le \tau_1 \le 1\} \in P. \nonumber
\end{eqnarray}

Define an integral, over this 4-path, of the 4-form $K$  determined by the antisymmetric tensor $K_{\alpha\beta\gamma\delta}$: 
\begin{eqnarray}
	I_K(w) 
&=& 
4! \int_0^1 d\tau_1 
\int_0^1 d\tau_2 
\int_0^1 d\tau_3 
\int_0^1 d\tau_4 \nonumber\\
&\times&
\partderiv{\xi^\alpha}{\tau_1}\cdot 
\partderiv{\xi^\beta}{\tau_2}\cdot 
\partderiv{\xi^\gamma}{\tau_3}\cdot 
\partderiv{\xi^\delta}{\tau_4}\cdot \nonumber\\
&\times& 
K_{\alpha\beta\gamma\delta}(p_{\tau_1}(\tau_2,\tau_3,\tau_4)) .
\label{CovarFourInt}
\end{eqnarray}
The integral is also additive: $I_K(ww') = I_K(w) + I_K(w')$. 

The (generalized) Stokes theorem is the equality, under certain conditions, of the 3-dimensional and 4-dimensional integrals (\ref{CovarThreeInt}) and (\ref{CovarFourInt}). They turn out to be equal if the following two conditions are fulfilled: (i)~$K$ is the covariant external derivative of 3-form $H$ (let us denote this as $K = DH$), and (ii)~the 3-path $v$ is a (generalized) boundary of the 4-path $w$ (denote this relation as $v = {\cal D}w$). In this case  
\begin{equation}
I_H({\cal D}w) 	 = I_{DH}(w) 
\label{GravThreeFourStokes}
\end{equation} 

The statement (i) means that form $K$ is determined by tensor $K_{\alpha\beta\gamma\delta} = \nabla_{[\alpha} H_{\beta\gamma\delta]}$ i.e. by the lifting of the usual covariant external derivative $\nabla_{[\mu} H_{\nu\kappa\lambda]}(x)$. 

The statement (ii) will be defined here as follows:\footnote{Other variants of this condition also leading to the equality may also be considered.} 1)~$v = \Delta w$, i.e. the 3-path $v$ is equal to the shift along the 4-path $w$, and 2)~all 3-paths of the family $v(\tau_4)$, $\tau_4\in [0,1]$ (including $v = v(1)$), are 3-loops, i.e. closed 3-paths, $v(\tau_4)\in\widehat L^{(3)}\subset\widehat P^{(3)}$. If we choose for simplicity the initial 3-path $v(0)$ to be a null hypersurface in some point, i.e. take $v(0) = 1\in\widehat L^{(3)}$, then the first condition looks simpler: $v = v(1)$ or 
\begin{equation}
	\widetilde\xi(\tau_1, \tau_2, \tau_3) = \xi(\tau_1, \tau_2, \tau_3, 1).
\end{equation}
 
The equality (\ref{GravThreeFourStokes}) is a gravitational covariant analogue of Stokes theorem connecting the 3-dimensional and 4-dimensional integrals. It may be proven in two steps: first, with the help of additivity of the integrals (\ref{CovarThreeInt}) and (\ref{CovarFourInt}), the equality (\ref{GravThreeFourStokes}) is reduced to the same equality but for infinitesimal $w$; second, the latter is directly verified (in the corresponding order) for an infinitesimal 4-dimensional cube. Decomposition of the finite integral (\ref{CovarFourInt}) in the sum of the infinitesimal integrals over infinitesimal 4-cubes is performed with the help of the structure (\ref{four-path}) of the 4-path $w$ as a framework. 

\section{Conservation of material energy-momentum in GR}
\label{sec:InterpretEnergyMomInteg}

The form $H$ in the Stokes theorem (\ref{GravThreeFourStokes}) may have additional indices, and covariant derivatives in the expression for the external derivative $DH$ should be calculated with all the indices taken into account, including also additional indices. Therefore, we may apply theorem (\ref{GravThreeFourStokes}) for the four 3-forms expressed in terms of EMT as  $H^{(\alpha')}_{\beta\gamma\delta} = T^{\alpha'\alpha} \,\epsilon_{\alpha\beta\gamma\delta}$.\footnote{$\nabla_{[\alpha} H^{(\alpha')}_{\beta\gamma\delta]}$ is defined in this case by lifting $\nabla_{[\mu} H^{(\mu')}_{\nu\lambda\kappa]}(x)$.} Because of the covariant derivative of EMT being zero, $\nabla_\alpha T^{\alpha'\alpha} = 0$, we obtain that $DH^{(\alpha)} = 0$ for all $\alpha$. Therefore, integral $I_{H^{(\alpha)}}(v)$ defined by Eq.~(\ref{CovarThreeInt}) is zero each time when the 3-path $v = {\cal D} w$ for some 4-path $w$: 
\begin{eqnarray}
I_{H^{(\alpha')}}({\cal D} w) = 
	\int_0^1 d\tau_1 \int_0^1 d\tau_2 \int_0^1 d\tau_3 
	 \nonumber\\	
	\partderiv{\widetilde\xi^\alpha}{\tau_1} \cdot 
	\partderiv{\widetilde\xi^\beta}{\tau_2} \cdot 
	\partderiv{\widetilde\xi^\gamma}{\tau_3} \; 
	 \epsilon_{\alpha''\alpha\beta\gamma}\,
T^{\alpha'\alpha''}(p_{\tau_1}(\tau_2,\tau_3))
	= 0. 
	\label{CovarEMTInt}
\end{eqnarray}

This is a \emph{covariant integral form of the conservation law for energy-momentum of matter}. It has an unusual feature: integral (\ref{CovarEMTInt}) is  expressed in terms of curves in the standard Minkowski space $\calM$ connected with the physical space-time $\calX$  in an non-holonomic way. 

One more peculiarity of the conservation law (\ref{CovarEMTInt}) is that it is written in the form of an operator equality. Explicit dependence of the integrand on local frames $n\in\calN$ arises only when the operator (\ref{CovarEMTInt}) acts on a function $\varphi(n)$ determined in $\calN$. If this function is concentrated in a small vicinity of some local frame $n_0$, then each path $p\in P$ in the integrand of (\ref{CovarEMTInt}) corresponds in fact to the frame $n = n_0\, p$ (which in turn determines the point $x\in\calX$ such that $n\in \calB_x$). In this case the properties of integral (\ref{CovarEMTInt}) become more transparent and can be compared with the analogous integral in the flat physical space-time (where there is no gravity). 

Both general structure and interpretation of integral (\ref{CovarEMTInt}) are just the same as for energy-momentum conservation in flat space-time (see Sect.~\ref{sec:Intro} after Eq.~(\ref{GaussFourDimNoncovar})). For the appropriate choice of a 3-path (ordered hypersurface) $v = {\cal D} w$ it may be divided in 3 parts: the bottom and top corresponding the 3-volume in the past and future and the side hypersurface corresponding to the surface of this volume in the intermediate times. If the integral over the side hypersurface is zero, then the energy-momentum in the past volume is equal to the energy-momentum in the future volume. 

This is the integral energy-momentum conservation law but formulated in terms of 3-paths (ordered volumes) instead of usual volumes. However, it can easily be seen that the integral over an ordered space-like volume in this construction does not depend of its ordering. To show this, we may form a 3-path $v = {\cal D} w$ with the null side hypersurface and bottom and top differing only by ordering (therefore covering the same volume). Then it follows from the conservation law that the integrals over bottom and top are equal, hence the volume integral does not depend on the ordering. 

\begin{figure}[h]
	\begin{center}
\parbox{3.5cm}{
\scalebox{0.40}
%\scalebox{0.25}
{\includegraphics{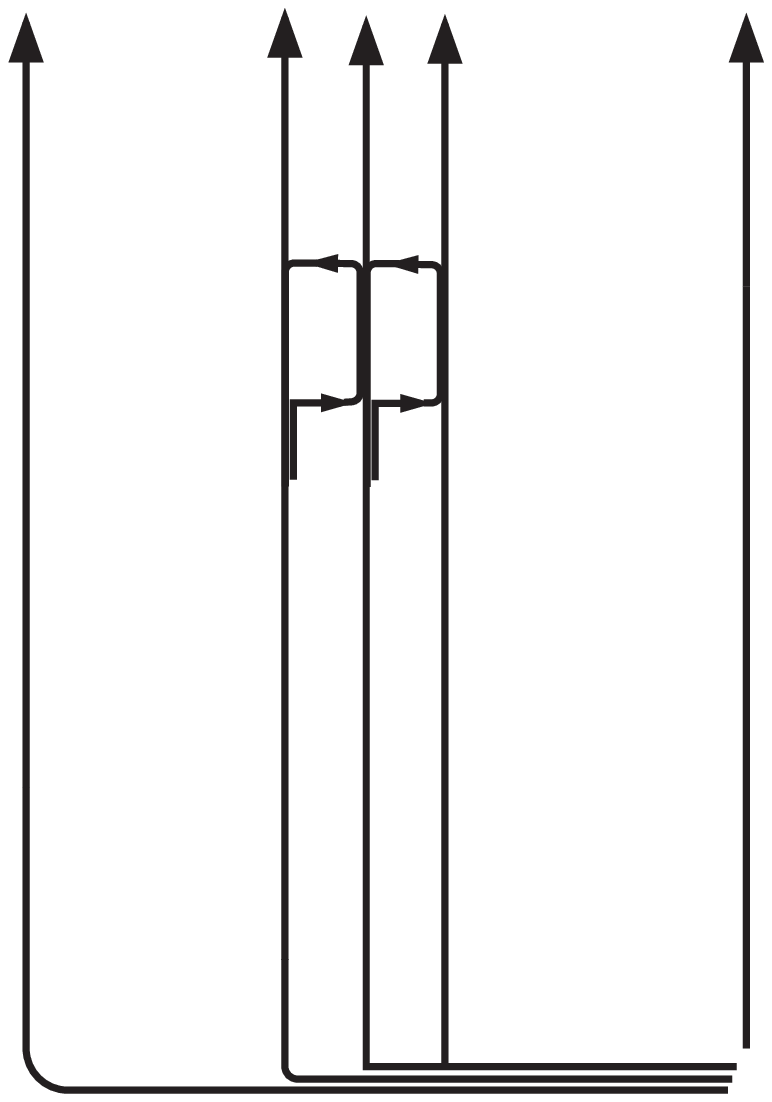}}
%\\Solid volume in ${\cal M}$
}
\quad \quad
\parbox{3.5cm}{
\scalebox{0.44}
%\scalebox{0.27}
{\includegraphics[viewport = 0 30 300 260]{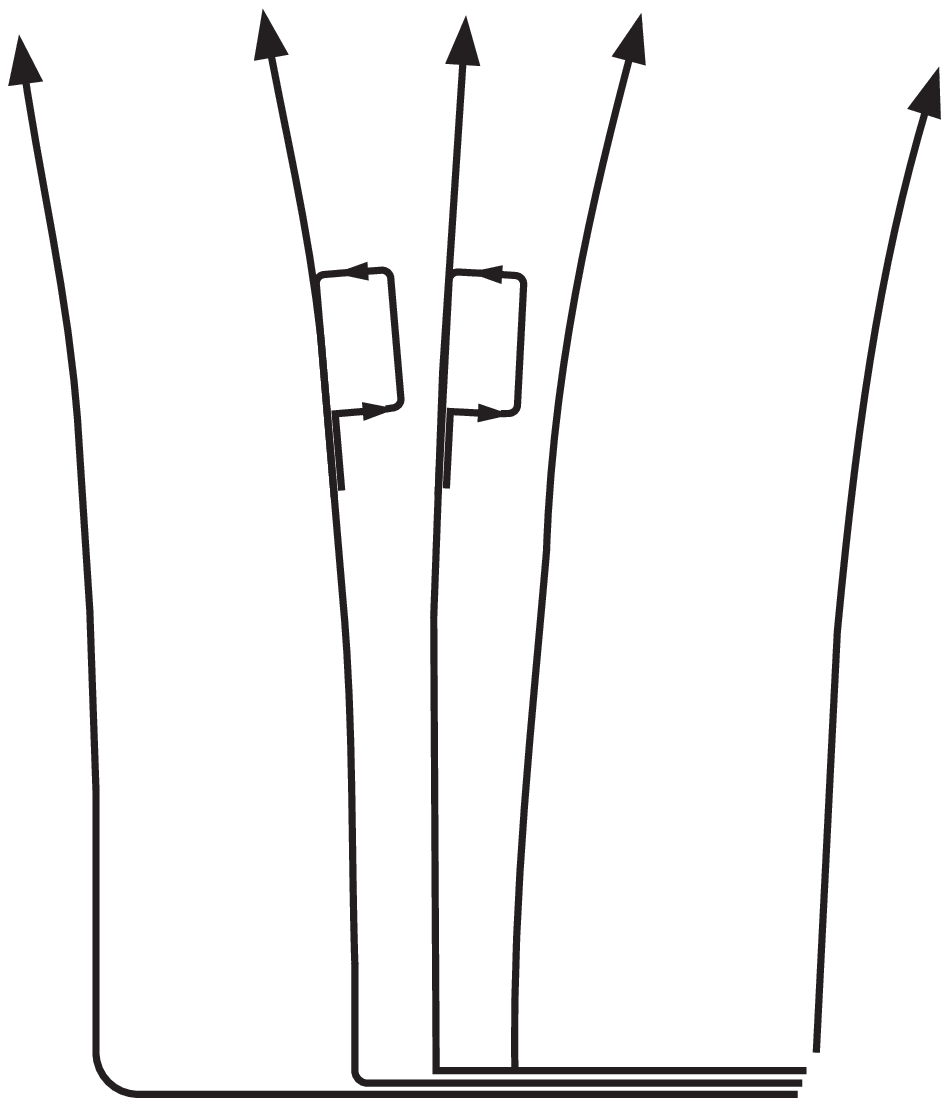}}
%\\[5mm] Dispersed volume in ${\cal X}$
	}
\end{center}
	\caption{\label{Fig1}The dispersed volume in ${\cal X}$ (right) corresponding to a solid volume in ${\cal M}$ (left)}
	\label{fig:dd221750}
\end{figure}

There is one more peculiarity of the conservation law (\ref{CovarEMTInt}) as compared with the flat-space-time one. The geometric elements (past and future volumes and side hypersurface) are described in (\ref{CovarEMTInt}) in terms of the `standard' Minkowski space-time $\calM$ which is connected with the physical space-time $\calX$ in an non-holonomic way. In particular, a continuous (solid) volume in $\calM$ corresponds in general case to a `dispersed volume' in $\calX$. 
Such a `volume' consists of distinct infinitesimal volume elements with infinitesimal spaces between them (or quite the reverse, infinitesimally intersecting with each other). This is illustrated in Fig.~\ref{Fig1} where the two-dimensional analogy is presented. An example of parallel paths in $\calM$ is drawn in the left part of the figure. In $\calX$ (right part) these paths correspond to a family of geodesics which are parallel at the beginning but diverge later. The small volumes filling the spaces between the parallel paths in $\calM$, correspond in $\calX$ to small volumes of the same size, but now they do not fill the spaces between the geodesics. 

Eq.~(\ref{GravThreeFourStokes}) describes the conservation of the energy-momentum of matter, without energy-momentum of the gravitational field. In this sense it extends the \emph{equivalence principle} from the law of motion to the energy-momentum conservation law: energy-momentum of matter alone is conserved, gravitational field is excluded from this conservation law. This cannot be done in terms of usual coordinate frames, but is naturally described by paths in subsidiary Minkowski space: elements of a dispersed volume are then brought together in the solid volume. 

If the paths in the 4-integral (\ref{CovarFourInt}) are direct lines (as in Fig.~\ref{Fig1}, but not necessarily parallel) describing the parts of the matter moving along geodesics, then the conservation law is interpreted as the conservation of the rest energy of matter. 

\section{Conclusion}
\label{sec:Conclusion}

It is commonly believed that no integral conservation law follows from the fact that the covariant derivative of the energy-momentum tensor (EMT) of matter is zero in General Relativity (\ref{CovarDiffConserv}). We showed above that this is wrong. Nullification of covariant derivative of EMT leads to the integral conservation law for energy-momentum of matter (without energy-momentum of gravitational field). This may be interpreted as an \textit{equivalence principle for energy}. 

Mathematically this conservation law is expressed in terms of the standard Minkowski space-time $\calM$ connected with the physical space-time $\calX$ in a non-holonomic way. Path groups provide an adequate mathematical formalism for dealing with $\calM$ and $\calX$. Fiber bundle $\calN$ of orthonormal local frames (having $\calX$ as its base) is essentially used in this formalism. Previously the equivalence principle (claiming that the material points move along geodesics) as well as its quantum generalization were formulated in terms of `standard' Minkowskian space-time $\calM$ and group $P(\calM)$ of paths in $\calM$ \cite{PGequivPrinciple}. 

One of the consequences of the non-holonomic connection between $\calM$ and $\calX$ is that a solid volume in $\calM$ (appearing in the discussed conservation law) corresponds to a dispersed volume in $\calX$. This feature is the reason why the energy conservation law formulated above in terms of the paths in the `standard' space-time $\calM$ cannot be expressed directly in terms of usual coordinates in the physical space-time $\calX$. The physical interpretation of this unusual property of energy in its relation to volume may be connected with the phenomenon of dark matter and serve as an explanation of this phenomenon. This will be discussed in a separate paper. 

\vskip 0.5cm
\centerline{\bf ACKNOWLEDGEMENTS}

The author is obliged to Horst von Borzeszkowski for fruitful discussions. The work was supported in part by the Russian Foundation of Basic Research, grant 02-01-00534a.

\newpage

\end{document}